\documentclass[prd,twocolumn,nofootinbib,notitlepage,10pt,aps]{revtex4-1}
\usepackage[utf8]{inputenc}
\usepackage{cancel}
\usepackage{xcolor}
\usepackage{latexsym}
\usepackage{amsmath}
\usepackage{amssymb}
\usepackage{bbm}

\usepackage{hyperref}
\hypersetup{
    colorlinks,
    linkcolor={violet},
    citecolor={teal},
    urlcolor={teal}
}
\usepackage{orcidlink}

\usepackage{ulem}
\usepackage{pdfsync}
\usepackage{epsfig}
\usepackage{epstopdf}
\usepackage{subfigure}
\usepackage{color}
\usepackage{comment}
\usepackage{slashed}
\usepackage{physics}

\usepackage{tensor}

\def\c{\chi}
\def\d{\delta}
\def\e{\epsilon}
\def\eps{\epsilon}
\def\g{\gamma}
\def\k{\kappa}
\def\l{\lambda}
\def\m{\mu}
\def\n{\nu}

\def\r{\rho}
\def\s{\sigma}
\def\t{\tau}

\def\z{\zeta}

\def\tns{\tensor}

\def\beq{\begin{equation}}
\def\eeq{\end{equation}}
\def\baq{\begin{eqnarray}}
\def\eaq{\end{eqnarray}}

\newcommand{\be}{\begin{equation}} % only untightened
\newcommand{\ee}{\end{equation}}
\newcommand{\bea}{\begin{eqnarray}} % only untightened
\newcommand{\eea}{\end{eqnarray}}

\newcommand{\bmp}{\noindent\begin{minipage}{16cm}}
\newcommand{\emp}{\end{minipage}\vskip 7mm} % 7mm untightened
\def\lsim{\mathrel{\raise.3ex\hbox{$<$\kern-.75em\lower1ex\hbox{$\sim$}}}}
\def\gsim{\mathrel{\raise.3ex\hbox{$>$\kern-.75em\lower1ex\hbox{$\sim$}}}}

\newcommand{\intron}[1]{}%{#1}

\begin{document}

\title{Reheating in geometric Weyl-invariant Einstein–Cartan gravity}

\author{Ioannis D. Gialamas\orcidlink{0000-0002-2957-5276}}
\email{ioannis.gialamas@kbfi.ee}

\affiliation{Laboratory of High Energy and Computational Physics, National Institute of Chemical Physics and Biophysics, 
R\"avala 10, 10143 Tallinn, Estonia}

\begin{abstract}

We study Weyl-invariant purely gravitational theories formulated within the Einstein–Cartan framework. In the Einstein-frame description, these models are dynamically equivalent to standard general relativity coupled to an axionlike pseudoscalar degree of freedom, which naturally drives a period of cosmic inflation. Without committing to a specific microscopic mechanism for reheating, we demonstrate that the postinflationary reheating dynamics play a crucial role in shaping the inflationary predictions. In particular, we show that assumptions about the reheating temperature and the equation-of-state parameter can significantly affect the predicted values of inflationary observables, highlighting the necessity of consistently incorporating reheating effects into the phenomenological analysis of inflationary models.

\end{abstract}

%%%%%%%%%%%%%%%%%%%%%%%%%%%%%%%%%%%%%%%%%%%%%%%%%%%%%%%%%%%%%%%%%%%%%%%%%%%%%%%%%%%%%%%%%%%%%%%%%%%%
% DOCUMENT
%
\maketitle
%%%%%%%%%%%%%%%%%%%%%%%%%%%%%%%%%%%%%%%%%%%%%%%%%%%%%%%%%%%%%%%%%%%%%%%%%%%%%%%%%%%%%%%%%%%%%%%%%%%%

%%%%%%%%%%%%%%%%%%%%%%%%%%%%%%%%%%%%%%%%%%%%%%%%%%%%%%%%%%%%%%%%%%%%%%%%%%%%%%%%%%%%%%%%%%%%%%%%%%%%
%%%%%%%%%%%%%%%%%%%%%%%%%%%%%%%%%%%%%%%%%%%%%%%%%%%%%%%%%%%%%%%%%%%%%%%%%%%%%%%%%%%%%%%%%%%%%%%%%%%%
%
\section{Introduction and framework}

Cosmic inflation~\cite{Kazanas:1980tx,Sato:1980yn,Guth:1980zm,Linde:1981mu} provides a compelling framework for describing the very early Universe, offering a simple explanation for the observed homogeneity and isotropy on large scales, as well as for the origin of primordial density perturbations~\cite{Starobinsky:1979ty,Mukhanov:1981xt,Hawking:1982cz,Starobinsky:1982ee,Guth:1982ec,Bardeen:1983qw}. A broad class of inflationary models successfully reproduces the current observational bounds~\cite{Planck:2018jri,BICEP:2021xfz,AtacamaCosmologyTelescope:2025nti} on the scalar spectral index $n_s$, the amplitude of primordial fluctuations $A_s$, and the tensor-to-scalar ratio $r$. As a result, many conceptually distinct scenarios remain phenomenologically viable, giving rise to a significant degeneracy among inflationary models.

This degeneracy motivates the exploration of inflation within more fundamental gravitational frameworks, where additional theoretical structure may lead to distinctive predictions. One such extension is Einstein--Cartan gravity, in which spacetime geometry is characterized not only by curvature but also by torsion\footnote{The Riemann and the torsion tensors are given by $
\tns{\mathcal{R}}{^\m_\nu_\k_\l}  = \partial_\k \tns{\Gamma}{^\m_\l_\n} - \partial_\l \tns{\Gamma}{^\m_\k_\n} + \tns{\Gamma}{^\m_\k_\s} \tns{\Gamma}{^\s_\l_\n} - \tns{\Gamma}{^\m_\l_\s}\tns{\Gamma}{^\s_\k_\n}$ and
$\tns{T}{^\k_\m_\n}  = \tns{\Gamma}{^\k_\m_\nu} - \tns{\Gamma}{^\k_\n_\m} $, respectively,
where $\Gamma^{\rho}{}_{\mu\nu}$ is a general affine connection. Additionally, the torsion can be decomposed into 
\begin{equation}
T_{\mu\nu\rho}=\frac{1}{3}\left(g_{\mu\rho} T_{\nu}-
g_{\mu\nu} T_{\rho}\right)+\frac{1}{6}\epsilon_{\mu\nu\rho\sigma}\hat{T}^{\sigma}+\tau_{\mu\n\rho} \,,
\end{equation}
where $T_{\mu} = \tns{T}{^\r_\m_\r}$ is the torsion vector, $\hat{T}^{\mu} = \epsilon^{\mu\nu\rho\sigma} T_{\nu\rho\sigma}$ is the axial vector, and $\tau_{\mu\nu\rho}$ denotes the purely tensorial part. }. In this framework, gravity is described by an affine connection that is metric compatible but not necessarily symmetric, allowing for a nonvanishing torsion tensor. As a consequence, new geometric invariants arise beyond those available in purely metric theories.

In the Einstein--Cartan framework, two independent curvature scalars can be constructed, namely the Ricci scalar $\mathcal{R}$ and the Holst invariant $\tilde{\mathcal{R}}$, the latter being a pseudoscalar quantity that violates parity. Imposing local Weyl invariance further constrains the form of the gravitational action, forbidding dimensionful couplings and restricting the dynamics to quadratic curvature combinations. As a result, the most general purely gravitational action that is Weyl invariant and constructed solely from curvature scalars, with overall mass dimension ${\rm mass}^4$ can schematically be written as
$\mathcal{S} \sim \gamma \mathcal{R}^2 + \delta \tilde{\mathcal{R}}^2 + \epsilon \mathcal{R}\tilde{\mathcal{R}}$,
where $\gamma$, $\delta$, and $\epsilon$ are dimensionless couplings.

Remarkably, despite the presence of higher-curvature terms, this theory admits a well-defined Einstein-frame description equivalent to the Einstein--Hilbert action supplemented by a single massive axionlike pseudoscalar field originated from the Holst invariant term in the initial action. The cosmological role of this pseudoscalar, both during inflation and beyond, has been extensively explored in various theoretical frameworks~\cite{Langvik:2020nrs,Shaposhnikov:2020gts,Piani:2022gon,Pradisi:2022nmh,Salvio:2022suk,Gialamas:2022xtt,Piani:2023aof,Gialamas:2023emn,DiMarco:2023ncs,He:2024wqv,Racioppi:2024zva,Inagaki:2024ltt,Gialamas:2024iyu,Racioppi:2024pno,Gialamas:2024uar,Karananas:2025xcv,Katsoulas:2025mcu,Gialamas:2025ciw,Salvio:2025izr,Gialamas:2025thp,Iosifidis:2025mcb,Karananas:2025fas,Karananas:2025qsm,Katsoulas:2025srh,Wang:2025cpp,Racioppi:2025igu}.
The role of the parity-odd term $\sim\mathcal{R}\tilde{\mathcal{R}}$ is particularly crucial. In its absence, the scalar potential grows exponentially and fails to support slow-roll inflation compatible with observations. Once the parity-breaking interaction is switched on, an inflationary plateau develops between the origin and the asymptotic exponential region of the potential, thereby rendering the model fully consistent with observations~\cite{Planck:2018jri,BICEP:2021xfz,AtacamaCosmologyTelescope:2025nti}.

While inflation determines the initial conditions for the hot big bang, it must be followed by a reheating phase~\cite{Liddle:2003as,Dodelson:2003vq,Martin:2010kz,Dai:2014jja,Munoz:2014eqa,Cook:2015vqa,Lozanov:2017hjm} during which the energy stored in the inflaton field is transferred to the ordinary Standard Model (SM) particles, resulting in a thermal radiation bath. Although the microphysics of reheating remains poorly understood, its macroscopic properties play a central role in connecting inflationary dynamics to late-time observables. In particular, the reheating temperature and the effective equation-of-state parameter during that epoch uniquely determine the relation between the number of $e$-folds $N_\star$ at horizon crossing and the values of observable quantities such as $n_s$ and $r$ (see, for example,~\cite{Gialamas:2019nly,Lahanas:2022mng} for explicit demonstrations of how reheating modifies the predictions of inflationary models with quadratic curvature terms).

In this paper, we investigate inflation and reheating in the Weyl-invariant model introduced above. We analyze how parity violation controls the shape of the inflationary potential and drives the predictions of the model toward those of the  Starobinsky model~\cite{Starobinsky:1980te}. We then study the impact of noninstantaneous reheating on inflationary predictions, treating the reheating phase in a phenomenological and model-independent manner. Our results demonstrate that reheating effects are essential for a consistent confrontation of these theories with current cosmological data and will play an increasingly important role as future experiments further tighten constraints on inflationary observables.

\section{The model}
In the Einstein-Cartan framework, a consistent gravitational theory must be constructed from an action containing at most quadratic combinations of curvature and torsion to avoid the appearance of higher-derivative terms. The requirement of Weyl invariance, which forbids dimensionful couplings, excludes linear contributions entirely. Consequently, only terms quadratic in the curvature are permitted. It is well known that generic quadratic curvature invariants can introduce additional, unwanted degrees of freedom, such as unhealthy ghosts~\cite{Stelle:1976gc}. Remarkably, these pathologies are absent for the special case of terms quadratic in the scalar curvature.

Focusing on a purely gravitational and Weyl-invariant action within the Einstein-Cartan framework, we consider the action
\begin{equation}
    \mathcal{S} = \int {\rm d}^4x \sqrt{-g} \left(\frac{\gamma}{4} \mathcal{R}^2 +\frac{\delta}{4} \tilde{\mathcal{R}}^2 +\frac{\epsilon}{2} \mathcal{R} \tilde{\mathcal{R}} \right)\,,
\label{eq:action_1}
\end{equation}
where $\mathcal{R}$ denotes the standard Ricci scalar  and $\tilde{\mathcal{R}}$ is the pseudoscalar Holst invariant given by
\begin{equation}
    \mathcal{R} =\tns{\mathcal{R}}{^\m^\n_\m_\n} \qquad \text{and} \qquad \tilde{\mathcal{R}} = \epsilon^{\mu\nu\kappa\lambda}\mathcal{R}_{ \mu\nu\kappa\lambda}\,.
\end{equation}
As already discussed, in the Einstein-Cartan framework, additional scalar terms such as $\mathcal{R}\phi^2$, $\tilde{\mathcal{R}}\phi^2$, $\hat{T}^\mu \hat{T}_\mu \phi^2$, etc., have the same mass dimension and could, in principle, be included in the action. We omit them, however, since they violate the assumption of a purely gravitational framework on which our analysis is based.

The action~\eqref{eq:action_1} admits an equivalent form in terms of two auxiliary scalar fields, $\chi$ and $\zeta$, and can be written as~\cite{Gialamas:2024iyu}
\begin{equation}
    \mathcal{S} = \int {\rm d}^4x \sqrt{-g} \left(\frac{\g\c}{2}\mathcal{R}+\frac{\d\z}{2}\tilde{\mathcal{R}} - \frac{\g\d (\g\c^2+\d\z^2-2\e\c\z)}{4(\g\d-\e^2)} \right)\,.
\end{equation}
As will be shown below, although two auxiliary scalar fields are introduced at this stage, the resulting theory contains only a single propagating axionlike, pseudoscalar degree of freedom. Furthermore, Weyl invariance of the action allows us to fix the gauge, which we choose as $\chi = M_P^2/\gamma$, where $M_P$ denotes the reduced Planck mass.

In order to obtain a metric equivalent formulation, we reexpress the curvature scalars in terms of the Levi-Civita connection and the components of torsion. Explicitly, one finds
\begin{equation}
{\cal{R}}=R[g]+2D_{\m}T^{\m}-\frac{2}{3}T_{ \m}T^{ \mu}+\frac{1}{24}\hat{T}^{\mu}\hat{T}_{\m}+\frac12\t_{ \m\n\r}\t^{ \m\n\r}\,
\end{equation}
and
\begin{equation}
\tilde{\mathcal{R}}=-D_{ \mu}\hat{T}^{ \mu}+\frac{2}{3}\hat{T}_{ \m}T^{ \m}+\frac{1}{2}\e^{ \m\n\r\s}\t_{ \lambda\mu\nu}\tns{\tau}{^\lambda_\rho_\sigma}\,,
\end{equation}
respectively.
Here $R[g]$ denotes the usual Ricci scalar constructed from the metric $g_{\mu\nu}$, while $D_{ \mu}$ is the covariant derivative associated with the Levi-Civita connection.

Substituting these expressions into the action and varying with respect to the torsion components $T_\m$, $\hat{T}_\m$, and $\t_{\m\n\rho}$, we obtain purely algebraic equations of motion for the torsion fields. Solving these equations and substituting the solutions back into the action, while imposing the gauge condition $\chi = M_P^2/\gamma$ and suitable field redefinition, yields the on-shell action
\begin{equation}
\mathcal{S}=\int {\rm d}^4x\sqrt{-g} \left[\frac{M_P^2}{2}R -\frac12(\partial_\m \phi)^2- V(\phi)-\frac{M_P^4}{4\g}\right]\,.
\label{eq:action_2}
\end{equation}
The resulting theory is thus equivalent to Einstein's gravity minimally coupled to a single canonical pseudoscalar field $\phi$, with a scalar potential given by\footnote{In action~\eqref{eq:action_2}, we also note the presence of a cosmological constant term $\propto \g^{-1}$. This should not be interpreted as the only contribution to the cosmological constant. Both perturbative and nonperturbative quantum corrections are expected to contribute to this term, potentially modifying it to yield the observed value. For example, the contribution from the SM, is generically expected to be $\mathcal{O}(M_W^4)$.}
\begin{equation}
V(\phi) = V_0 \left(4\theta +\sinh\left[\sqrt{\frac{2}{3}}\frac{\phi}{M_P}-{\rm arcsinh} [4\theta] \right] \right)^2 \,,
\label{eq:potential}
\end{equation}
where we have introduced the dimensionless parameter $\theta = \e/(2\g)$ and the overall normalization $V_0 =M_P^4/(16(\d-4\g\theta^2))$. The structure of the potential is fully determined by the underlying gravitational sector and encodes the effects of both parity even and parity odd curvature invariants. The scalar representation of this purely gravitational theory in the presence of the mixed $\mathcal{R}\tilde{\mathcal{R}}$ term was first derived in~\cite{Gialamas:2024iyu}, and its inflationary predictions have been discussed in~\cite{Karananas:2025xcv}. In this paper, we focus on the reheating dynamics and investigate their influence on the resulting inflationary observables.

Remarkably, exactly the same scalar potential has appeared in several seemingly different gravitational constructions that are parity odd, but lack Weyl invariance. It arises in purely gravitational theories built from actions of the schematic form $\mathcal{S} \sim \mathcal{R}+\mathcal{R}^2+\tilde{\mathcal{R}}+\tilde{\mathcal{R}}^2$, previously studied in~\cite{Pradisi:2022nmh,Salvio:2022suk,Gialamas:2022xtt}. An identical potential is also realized in some limits of nonminimally coupled models, which share common features with $\xi$-attractor scenarios~\cite{Kallosh:2013tua,Galante:2014ifa}, as discussed in~\cite{Racioppi:2024pno,Racioppi:2025igu}.

Another noteworthy feature of this model is that, in the large $\theta$ regime the scalar potential simplifies and approaches~\cite{He:2024wqv,Racioppi:2024pno,Katsoulas:2025srh}
\begin{equation}
V(\phi) \xrightarrow{\theta\gg1} 16\theta^2 V_0 \left(1-e^{-\sqrt{\frac{2}{3}}\frac{\phi}{M_P}} \right)^2 + \cdots,
\label{eq:pot_staro}
\end{equation}
reducing at leading order to the familiar form of the Starobinsky potential~\cite{Starobinsky:1980te}.

\section{Inflation and Reheating}
In this Section, we briefly analyze the inflationary predictions of the model, following the discussion in~\cite{Salvio:2022suk,Karananas:2025xcv}, and subsequently turn to the analysis of the reheating phase.

\subsection{Inflationary predictions}

Current observational bounds on inflationary observables, derived from the latest combined analyses of Planck, BICEP/Keck, and BAO data~\cite{Planck:2018jri,BICEP:2021xfz}, as well as from Planck, ACT, and DESI measurements~\cite{AtacamaCosmologyTelescope:2025nti}, provide a useful benchmark for assessing the viability of the model. We begin by considering the normalization of the scalar power spectrum generated during single-field slow-roll inflation. On super Hubble scales, the amplitude of curvature perturbations is given by
\begin{equation}
A_s^\star = \frac{H_\star^2}{8\pi^2 \eps_1^\star}\,,
\end{equation}
and is constrained at the pivot scale $k_\star = 0.05\,{\rm Mpc}^{-1}$ to the value $A_s^\star \simeq 2.1 \times 10^{-9}$~\cite{Planck:2018jri}. The remaining inflationary observables of interest, namely the tensor-to-scalar ratio, the scalar spectral index and its running, are expressed in terms of the Hubble-flow parameters as
\begin{align}
r = 16 \eps_1\,,  
\qquad &
n_s = 1 - 2 \eps_1 - \eps_2\,, \nonumber
\\[0.2cm]
{\rm d}n_s/ {\rm d}\ln k &= -2\eps_1\eps_2-\eps_2\eps_3\,.
\end{align}
Here $\eps_1$, $\eps_2$ and $\epsilon_3$ denote the first three Hubble-flow parameters, defined respectively as $\eps_1 \equiv - {\rm d}\ln H / {\rm d}N$,  $\eps_2 \equiv {\rm d}\ln \eps_1 / {\rm d}N$ and $\eps_3 \equiv {\rm d}\ln \eps_2 / {\rm d}N$, with ${\rm d}N = H\,{\rm d}t$. Due to the nontrivial structure of the potential in~\eqref{eq:potential}, analytic expressions for the inflationary observables are difficult to obtain. For this reason, we determine the background dynamics by numerically solving the equations of motion and compute the observables directly from the Hubble-flow functions, rather than relying on the standard potential slow-roll approximation.

The role of the parity violating term $\mathcal{R}\tilde{\mathcal{R}}$ in the action~\eqref{eq:action_1} is directly reflected in the structure of the scalar potential~\eqref{eq:potential} arising in the metric equivalent formulation~\eqref{eq:action_2}, through its dependence on the parameter $\theta$. In the absence of the parity odd term, $\theta$ vanishes identically and the resulting scalar potential exhibits a purely exponential behavior, rendering the model incompatible with successful slow-roll inflation and current observational constraints. 

By contrast, when the parity breaking term is present, i.e., $\theta$ is nonzero, the shape of the potential is significantly modified. As $\theta$ increases, an extended inflationary plateau develops between the origin and the asymptotic exponential regime. This plateau provides the necessary conditions for sustained slow-roll inflation. In the limit of large $\theta$, the flat region closely reproduces the characteristic Starobinsky plateau~\eqref{eq:pot_staro}, and the inflationary predictions of the model converge to those of Starobinsky inflation. 
Our numerical analysis indicates that this regime is effectively realized for $\theta \gtrsim 150$. 

This behavior is illustrated in Fig.~\ref{Fig:fig1}, where the gray solid curve shows the predictions obtained by varying the parameter $\theta$ over the range $\theta \in[15, 3000]$. The upper panel displays the resulting trajectories in the $(n_s,r)$ plane, while the lower panel shows the running of the scalar spectral index ${\rm d}n_s/{\rm d}\ln k$ as a function of $r$. As $\theta$ increases along the curve, the predictions smoothly approach the Starobinsky limit (black solid line). For smaller values of $\theta$, the inflationary plateau progressively disappears and the corresponding predictions move away from the region favored by current observations.

In the discussion above, we have assumed instantaneous reheating at the end of inflation, corresponding to a sudden transition to a radiation dominated phase of the Universe. As we shall see, relaxing this assumption can significantly affect the inflationary observables and modify the resulting predictions.

\subsection{Reheating}

In this Section, we turn to the postinflationary reheating phase and the parameters that characterize it. After inflation ends, the energy stored in the inflaton field must be redistributed to other particle species, initiating a sequence of interactions that eventually thermalize the Universe and generate a bath of SM particles. The specifics of this energy transfer, including the efficiency of particle production, possible nonperturbative effects such as preheating, and the duration of the reheating period, remain largely uncertain. These details can have a pronounced impact on the subsequent expansion history of the Universe and, in turn, on the precise values of inflationary observables. Understanding these dynamics is therefore crucial for connecting inflationary predictions with cosmological data.

\begin{figure}[t!]
    \centering
    \includegraphics[width=0.485\textwidth]{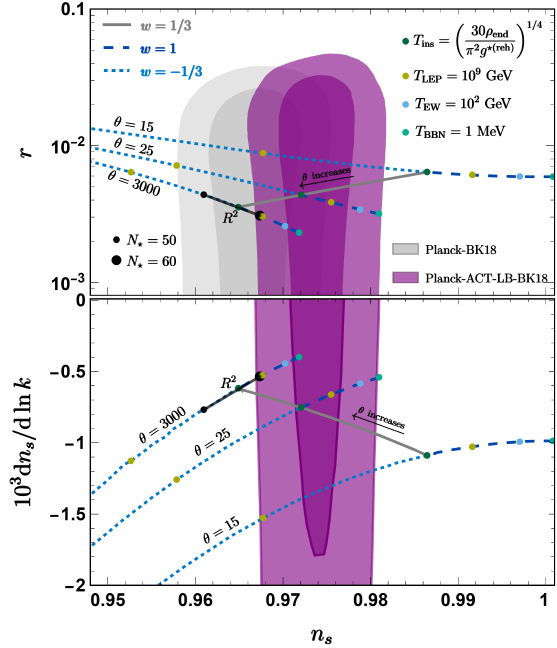}%
\caption{Tensor-to-scalar ratio $r$ versus scalar spectral index $n_s$ (upper panel) and running of the scalar spectral index ${\rm d}n_s/{\rm d}\ln k$ versus $n_s$ (lower panel), evaluated at the pivot scale $k_\star = 0.05\,{\rm Mpc}^{-1}$. The $1\sigma$ and $2\sigma$ confidence regions obtained from the latest combination of Planck, BICEP/Keck, and BAO data~\cite{Planck:2018jri,BICEP:2021xfz} and from Planck, ACT, and DESI data~\cite{AtacamaCosmologyTelescope:2025nti} are shown in gray and purple, respectively. The gray solid curve corresponds to the predictions of the model assuming instantaneous reheating, with the parameter $\theta$ varying in the range $\theta \in [15,3000]$. The dark blue dashed and light blue dashed curves show the predictions for a noninstantaneous reheating scenario with equation-of-state parameters $w=1$ and $w=-1/3$, respectively, for representative values of the parameter $\theta = 15, 25, 3000$. Moving away from the gray solid curve along the dashed lines, the reheating temperature decreases. At the same time, the number of $e$-folds increases when moving to the right and decreases when moving to the left. Colored dots indicate representative reheating temperatures associated with big bang nucleosynthesis (BBN), electroweak, and leptogenesis scales, as well as the maximum temperature obtained in the instantaneous reheating limit. For comparison, the black solid curve shows the predictions of the Starobinsky model~\cite{Starobinsky:1980te}.
} 
   \label{Fig:fig1}
\end{figure}
In the following, we adopt a model-independent approach and do not propose a particular reheating mechanism. Instead, the process is described in terms of a small set of phenomenological parameters. First, we characterize the duration of the reheating period by the number of e-folds 
\begin{equation}
N_{\rm reh} \equiv \ln \frac{a_{\rm reh}}{a_{\rm end}}\,,
\end{equation}
where $a_{\rm end}$ and $a_{\rm reh}$ denote the scale factors at the end of inflation and at the completion of reheating, respectively (in the case of instantaneous reheating, $a_{\rm end} = a_{\rm reh}$ and thus $N_{\rm reh} = 0$). The evolution of the total energy density during reheating, $\rho \propto a^{-3(1+\omega)}$, is captured by an effective equation-of-state parameter, $\omega$, defined as the average
\begin{equation}
\omega \equiv \frac{1}{N_{\rm reh} - N_{\rm end}} \int_{N_{\rm end}}^{N_{\rm reh}} \omega(N) \, {\rm d}N\,,
\end{equation}
where $N_{\rm end}$ denotes the number of $e$-folds at the end of inflation, and $\omega(N) = P/\rho$ relates the pressure $P$ to the energy density $\rho$. This definition provides a simple yet effective way to characterize the average behavior of the energy content of the Universe during the reheating period. At the end of inflation, the equation-of-state parameter is $\omega(N_{\rm end}) = -1/3$, while at the end of reheating it approaches $\omega(N_{\rm reh}) = 1/3$, corresponding to the transition into a radiation dominated Universe. In the absence of a detailed reheating model, the precise evolution of $\omega$ remains largely uncertain. For this reason, we treat $\omega$ as a free parameter, exploring values within a physically reasonable range, $-1/3\leq w\leq 1$, to capture the variety of possible reheating scenarios.

Finally, we introduce the reheating temperature $T_{\rm reh}$, which corresponds to the temperature of the thermalized plasma at the conclusion of the reheating period. This parametrization allows us to capture the essential features of reheating without specifying the microscopic details of particle production or thermalization.

To incorporate the effects of the reheating phase into the evolution of the Universe, one must compute the number of $e$-folds between the moment when the pivot scale, $k_\star$, crosses the horizon and the end of inflation, $N_\star \equiv \ln (a_{\rm end}/a_\star)$, explicitly taking into account the subsequent reheating phase.  The horizon crossing condition is given by
$a_\star H_\star = k_\star$\,,
with $H_\star$ being the Hubble parameter at horizon exit. 

This relation can be rewritten to connect the various stages of expansion as
\begin{equation}
H_\star = \frac{a_{\rm end}}{a_\star} \frac{a_{\rm reh}}{a_{\rm end}}\frac{a_0}{a_{\rm reh}}\frac{k_\star}{a_0}\,,
\label{eq:scales}
\end{equation}
where the subscript “0” denotes present-day values.

Using the fact that at the end of reheating, the total energy density is fully transferred into ultrarelativistic species, the energy density can be written as 
\begin{equation}
\rho_{\rm reh} = \frac{\pi^2}{30}g^{\star({\rm reh})}T_{\rm reh}^4\,,
\label{eq:rho_reh}
\end{equation}
where $g^{\star({\rm reh})}$ is the energy density degrees of freedom.
Now, using the conservation of entropy along with Eqs.~\eqref{eq:scales} and~\eqref{eq:rho_reh} allows one to consistently relate the reheating temperature to the expansion history of the Universe and, in turn, to express the number of $e$-folds between the horizon exit of the pivot scale, $k_\star$, and the end of inflation as\footnote{The numerical value of the first term in Eq.~\eqref{eq:Nstar} evaluates to $\ln \left[\left(\frac{\pi^2}{30}\right)^{1/4} \frac{{g_s^{\star(0)}}^{1/3}}{\sqrt{3}}\frac{T_0}{H_0} \right] =66.89$. In obtaining this result, we have used the present-day temperature $T_0 = 2.35 \times 10^{-4}\,{\rm eV}$, the reduced Hubble constant $h = 0.676$, and the entropy degrees of freedom today $g_s^{\star(0)} = 43/11$. For the reheating phase, we assume $g_s^{\star({\rm reh})} = g^{\star({\rm reh})} = 106.75$, corresponding to the SM particle content at temperatures above $1\,{\rm TeV}$. Note also that a term $\dfrac14 \ln\left[g^{\star({\rm reh})}/g_s^{\star({\rm reh})}\right]$ on the rhs of~\eqref{eq:Nstar} is very small and has been omitted.}
\begin{align}
\label{eq:Nstar}
    N_\star =& \ln \left[\left(\frac{\pi^2}{30}\right)^{1/4} \frac{{g_s^{\star(0)}}^{1/3}}{\sqrt{3}}\frac{T_0}{H_0} \right] -\ln\left[\frac{k_\star}{a_0 H_0}\right] 
\\
& - \frac{1}{12}\ln \left[g_s^{\star({\rm reh})}\right]+\frac14\ln \left[\frac{9H_\star^4}{\rho_{\rm end}}\right] + \frac{3w-1}{4}N_{\rm reh}\,, \nonumber
\end{align}
where the number of $e$-folds during reheating is given by
\begin{equation}
\hspace{-0.16cm}  N_{\rm reh} = -\frac{4}{3(1+w)} \bigg(\ln\left[\frac{T_{\rm reh}}{\rho_{\rm end}^{1/4}}\right] +\frac14 \ln \left[\frac{\pi^2 g^{\star({\rm reh})} }{30}\right] \bigg)\,,
\end{equation}
and the corresponding reheating temperature by
\begin{equation}
    T_{\rm reh}= \left(\frac{30\rho_{\rm end}}{\pi^2 g^{\star({\rm reh})}} \right)^{1/4} e^{-\frac{3(1+w)N_{\rm reh}}{4}}\,.
\label{eq:Treh}
\end{equation}
For $w \neq 1/3$, the reheating phase spans a nonzero number of $e$-folds, and therefore due to $w>-1$ the reheating temperature is bounded from above
\begin{equation}
    T_{\rm reh}\leq \left(\frac{30\rho_{\rm end}}{\pi^2 g^{\star({\rm reh})}} \right)^{1/4} \equiv T_{\rm ins}\,,
\end{equation}
where the right-hand side defines the instantaneous reheating temperature $T_{\rm ins}$, corresponding to the limiting case in which reheating occurs immediately at the end of inflation. 
Adopting $g^{\star({\rm reh})} = 106.75$, as appropriate for the SM particle content, one finds the general relation
\begin{equation}
T_{\rm ins} \simeq 0.41\, \rho_{\rm end}^{1/4}\,.
\end{equation}
The precise value of $T_{\rm ins}$ therefore depends on the energy density at the end of inflation, $\rho_{\rm end}$, which is model dependent.

The end of inflation is defined by the condition $\eps_1 = 1$, or equivalently by $\rho_{\rm end} + 3P_{\rm end} = 0$, implying $\rho_{\rm end} = 3V(\phi_{\rm end})/2$. In this way, $\rho_{\rm end}$ is fully determined by the value of the scalar field at the end of inflation. To estimate $\phi_{\rm end}$ analytically, we employ the potential slow-roll parameters
$\eps_V = \frac{M_P^2}{2}\left(\frac{\partial_\phi V}{V}\right)^2$ and
$\eta_V = M_P^2\frac{\partial_\phi^2 V}{V}$,
since, to a good approximation, the termination of inflation can be inferred from either $\eps_V = 1$ or $|\eta_V| = 1$.

\begin{figure}[h!]
    \centering
    \includegraphics[width=0.48\textwidth]{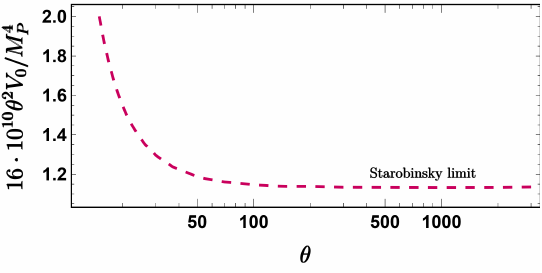}%
\caption{The normalized product $\theta^2 V_0$ as a function of $\theta$. In the large-$\theta$ regime, $\theta^2 V_0$ asymptotes to a constant, signalling the recovery of the Starobinsky limit (see Eq.~\eqref{eq:pot_staro}).}
   \label{Fig:fign}
\end{figure}

In the present model, a numerical analysis indicates that the condition $|\eta_V| = 1$ provides a remarkably accurate estimate of the end of inflation as determined by $\eps_1 = 1$. This behavior differs from the conclusions reported in~\cite{Karananas:2025xcv}. The resulting field value at the end of inflation is given by
\begin{align}
\frac{\phi_{\rm end}}{M_P} = &\sqrt{\frac{3}{2}}\bigg( {\rm arcsinh}[4\theta]
+\ln \bigg[\frac{1}{5}\bigg(4\theta - 2\sqrt{64\theta^2 - 5} \nonumber\\
&+\sqrt{272\theta^2 + 5 - 16\theta\sqrt{64\theta^2 - 5}}\bigg)\bigg] \bigg)\,.
\end{align}
In the large-$\theta$ limit this expression reduces to $\phi_{\rm end} \simeq (\sqrt{3}/2)\ln[5/3]\, M_P$, which should be compared with the more accurate value $\phi_{\rm end} = 0.615\, M_P$ obtained in the Starobinsky model from a numerical integration of the Einstein and Klein-Gordon equations.
Substituting this result into the definition of the instantaneous reheating temperature yields
\begin{equation}
\hspace{-0.12cm} T_{\rm ins} \simeq 0.29\, V_0^{1/4}
\left(8\theta\left(26\theta - 3\sqrt{64\theta^2 - 5}\right) - 5\right)^{1/4}\,.
\end{equation}
The parameter $V_0$ is itself $\theta$ dependent\footnote{As already mentioned, the parameter $V_0$ is given by $V_0=M_P^4/(16(\d-4\g\theta^2))$. Therefore, for a fixed value of $\theta$, the normalization of the scalar power spectrum can be adjusted by appropriately choosing the parameters $\g$ and $\d$ that multiply the parity-even quadratic curvature terms.} and can be determined by matching the amplitude of the scalar power spectrum, $A_s^\star \simeq 2.1 \times 10^{-9}$ at the pivot scale $k_\star = 0.05\,{\rm Mpc}^{-1}$. For each choice of $\theta$, this normalization condition fixes $V_0$ to a distinct value. The resulting $\theta$ dependence of $V_0$ is illustrated in Fig.~\ref{Fig:fign}.
\begin{figure*}[t!]
    \centering
    \includegraphics[width=\textwidth]{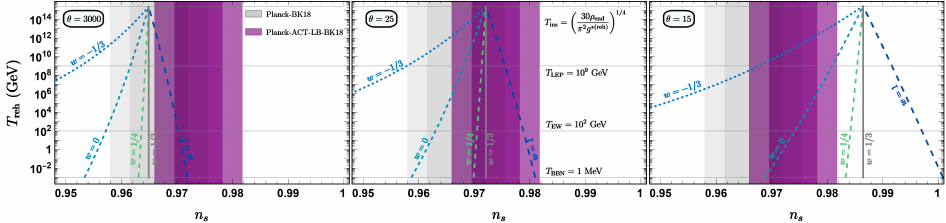}%
\caption{Reheating temperature $T_{\rm reh}$ as a function of the scalar spectral index $n_s$ for $\theta = 3000$ (left), $\theta = 25$ (middle), and $\theta = 15$ (right). Dashed curves correspond to noninstantaneous reheating scenarios with equation-of-state parameters $w = -1/3, 0, 1/4$, and $1$ (color coded as shown in the figure), while the instantaneous reheating case ($w = 1/3$) is represented by the solid vertical line. All curves intersect at the point where reheating occurs instantaneously.  Gray horizontal lines indicate the characteristic energy scales discussed in Fig.~\ref{Fig:fig1}. The observational confidence regions are the same as those shown in Fig.~\ref{Fig:fig1}.}
   \label{Fig:fig2}
\end{figure*}
In the large-$\theta$ regime one finds $V_0 \propto \theta^{-2}$, in agreement with expectations, since in this limit the potential smoothly approaches the Starobinsky form~\eqref{eq:pot_staro}. As a consequence, the combination $\theta^2 V_0$ asymptotes to a constant value, reflecting the recovery of the Starobinsky normalization. In this limit the instantaneous reheating temperature becomes independent of $\theta$. Substituting the asymptotic form of $V_0$ into the expression for $T_{\rm ins}$, one obtains $T_{\rm ins} \simeq 2.28 \times 10^{15}\,{\rm GeV}$, which coincides with the value expected in the Starobinsky model. Note that, in light of the latest data from ACT~\cite{AtacamaCosmologyTelescope:2025nti}, the original Starobinsky scenario is marginally incompatible, and several approaches have been proposed to reconcile it with observations, including detailed analyses of the reheating phase~\cite{Drees:2025ngb,Zharov:2025zjg,Antoniadis:2025pfa,Ellis:2025zrf} as well as extensions of the action with higher-curvature terms~\cite{Kim:2025dyi,Gialamas:2025ofz,Addazi:2025qra,Ketov:2025cqg} or the inclusion of radiative corrections~\cite{Ellis:2025bzi}.

As discussed above, the instantaneous reheating temperature represents an upper bound on the reheating scale and is realized only in the limiting case $N_{\rm reh}=0$ and $w=1/3$. At the same time, a robust lower bound applies, since the reheating temperature must remain above ${\cal O}(1\,{\rm MeV})$ in order not to spoil the successful predictions of BBN. In noninstantaneous reheating scenarios, the detailed dynamics of the reheating phase can significantly affect the value of $T_{\rm reh}$. In particular, the reheating temperature decreases exponentially with the number of reheating $e$-folds, $N_{\rm reh}$, allowing for values that can be many orders of magnitude smaller than the instantaneous limit.

The effect of a noninstantaneous reheating phase on the inflationary predictions is illustrated in Fig.~\ref{Fig:fig1}. There, the dark blue dashed and light blue dashed curves correspond to equation-of-state parameters $w=1$ and $w=-1/3$, respectively. Values of the equation-of-state parameter larger than $1/3$ lead to an increase in the number of inflationary $e$-folds, $N_\star$, and as a consequence the predictions in both panels of Fig.~\ref{Fig:fig1} are shifted toward larger values of the scalar spectral index $n_s$.

This effect is particularly pronounced in the Starobinsky limit, illustrated by the case $\theta=3000$ in Fig.~\ref{Fig:fig1}. In this regime, the combined Planck, ACT, and DESI data favor relatively stiff equations of state during reheating, with $w$ close to unity, together with low reheating temperatures approaching the BBN bound. Such a combination is required in order for the model predictions to fall within the allowed observational regions. This behavior is consistent with earlier studies of reheating in the Starobinsky model and closely related scenarios~\cite{Drees:2025ngb,Zharov:2025zjg}.

For smaller values of $\theta$, the situation is reversed. For instance, in the case $\theta=15$ shown in Fig.~\ref{Fig:fig1}, the observational data favor values of the equation-of-state parameter below $1/3$, which leads to a reduction of $n_s$ and brings the predictions into better agreement with the data. In contrast, for intermediate values of $\theta$ that are already compatible with observations in the instantaneous reheating scenario, no strong preference for a specific equation-of-state parameter emerges. Note that moving away from the gray solid curve (i.e. instantaneous reheating scenario) along the dashed lines, the reheating temperature decreases.
\begin{figure*}[t!]
    \centering
    \includegraphics[width=0.66\textwidth]{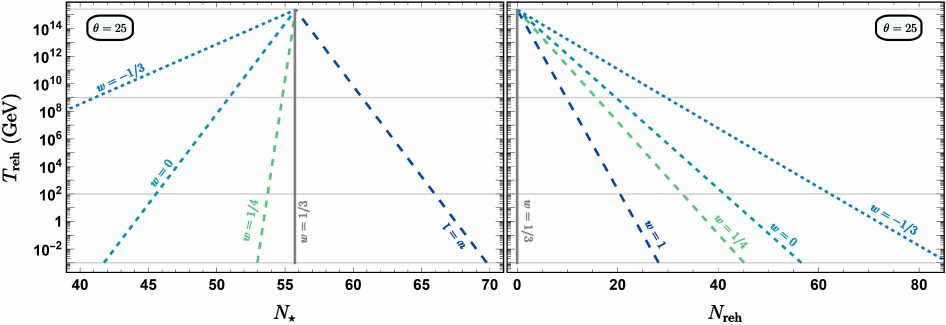}%
\caption{Reheating temperature $T_{\rm reh}$ as a function of the number of $e$-folds $N_\star$ (left) and $N_{\rm reh}$ (right), for $\theta = 25$ and equation-of-state parameters $w = -1/3, 0, 1/4, 1/3$, and $1$, as in Fig.~\ref{Fig:fig2}.}
   \label{Fig:fig3}
\end{figure*}

Further insight into the reheating dynamics is provided in Fig.~\ref{Fig:fig2}, where the reheating temperature is shown as a function of the scalar spectral index, $n_s$, for $\theta=3000$ (left panel), $\theta=25$ (middle panel), and $\theta=15$ (right panel). In the Starobinsky limit, i.e. $\theta=3000$, agreement with observational constraints requires $w>1/3$. For $w=1$, consistency with the $1\sigma$ and $2\sigma$ observational regions allows reheating temperatures in the range $1\,{\rm MeV}\lesssim T_{\rm reh}\lesssim 2\times10^{13}\,{\rm GeV}$. For $\theta=15$, a decrease of $n_s$ can be obtained, for instance, for $w=0$ (canonical reheating scenario) with $1\,{\rm MeV}\lesssim T_{\rm reh}\lesssim 7\times10^{9}\,{\rm GeV}$, or for $w=-1/3$ with $3\times10^{8}\,{\rm GeV}\lesssim T_{\rm reh}\lesssim 4\times10^{13}\,{\rm GeV}$. Finally, for $\theta=25$, a wide range of values for both $w$ and $T_{\rm reh}$ remains compatible with the data, reflecting a weaker sensitivity to the details of the reheating phase.
Finally, Fig.~\ref{Fig:fig3} displays the reheating temperature as a function of $N_\star$ (left panel) and $N_{\rm reh}$ (right panel) for several reheating scenarios with $\theta=25$. Both panels exhibit only a very weak dependence on the value of $\theta$, and the corresponding curves remain nearly unchanged for different choices of this parameter. This is because $T_{\rm reh}$ is an input parameter in Eq.~\eqref{eq:Nstar} and does not depend on $\theta$, while $N_\star$ exhibits only a mild logarithmic dependence on it. In contrast, the spectral index and the tensor-to-scalar ratio vary rapidly with $\theta$, as can be seen from its explicit appearance in the scalar potential~\eqref{eq:potential}.

\section{Discussion and Conclusions}

Inflationary cosmology encompasses a broad class of models that often yield nearly indistinguishable predictions for the limited set of inflationary observables currently accessible to observations. The reheating epoch offers an additional handle for lifting this degeneracy. Although the CMB does not directly probe the microphysics of reheating, the properties of this phase are intimately connected to inflationary predictions, as they shape the thermal history of the Universe immediately after inflation.

Despite the fact that the detailed dynamics of reheating remain largely unknown, it is possible to adopt physically motivated assumptions, such as bounding the average equation-of-state parameter during reheating within the range $-1/3 \leq w \leq 1$. These assumptions translate into nontrivial constraints on inflationary observables, including the scalar spectral index $n_s$ and the tensor-to-scalar ratio $r$, which can then be confronted with current CMB measurements.

In this work, we have explored a class of purely gravitational theories formulated within the Einstein–Cartan framework. The starting point is an action constructed from the two independent curvature scalars available in this setting, namely the Ricci scalar and the Holst invariant. Imposing Weyl invariance severely restricts the allowed operators and leads uniquely to quadratic curvature combinations of the form $\gamma \mathcal{R}^2$, $\d \tilde{\mathcal{R}}^2$, and $\e \mathcal{R}\tilde{\mathcal{R}}$. Unlike its metric counterpart based solely on $R^2$ gravity, the resulting theory is dynamically equivalent to Einstein gravity coupled to a massive axionlike pseudoscalar degree of freedom. Remarkably, this additional field naturally plays the role of the inflaton and gives rise to inflationary predictions fully consistent with current observational constraints.

A key ingredient of this construction is the parity-violating term $\mathcal{R}\tilde{\mathcal{R}}$. Its presence is crucial for the structure of the scalar potential in the Einstein-frame description. In the absence of this term, the potential exhibits a purely exponential behavior, rendering the model incompatible with observational data. By contrast, when the parity-odd term is included, an inflationary plateau emerges between the origin and the asymptotic exponential regime. In the limit of large values of the ratio $\theta=\e/(2\g)$, this plateau smoothly approaches that of the Starobinsky model.

We have further investigated the postinflationary reheating phase and its impact on the inflationary predictions of the model. Our analysis shows that reheating plays a decisive role in determining the allowed parameter space. In the Starobinsky limit, corresponding to large $\theta$, compatibility with current data favors reheating scenarios with equation-of-state parameters larger than $1/3$ and reheating temperatures ranging from values close to the BBN bound up to $T_{\rm reh}\sim 10^{13}\,{\rm GeV}$. For smaller values of $\theta$, such as $\theta=15$, the situation is reversed: observational consistency requires softer reheating equations of state, $w<1/3$, accompanied by a corresponding reduction of the scalar spectral index and reheating temperatures lower than $\sim 10^{13}\,{\rm GeV}$. Intermediate values of $\theta$ exhibit a milder dependence on the reheating dynamics and allow for a broader range of reheating temperatures and equations of state.

While in this work we treat reheating phenomenologically through an effective equation-of-state parameter $w$, it is useful to briefly comment on the types of mechanisms that could realize such a phase. After the end of inflation the inflaton typically undergoes coherent oscillations around the minimum of its potential and transfers its energy to lighter degrees of freedom. If the minimum is approximately quadratic, these oscillations correspond on average to an equation of state $w\simeq 0$. The subsequent decay of the inflaton, either through perturbative processes or through nonperturbative effects such as parametric resonance during preheating, then converts the energy density into radiation and eventually leads to the radiation dominated era with $w \to 1/3$. The range of $w$ values considered in our analysis should therefore be interpreted as an effective description of the intermediate dynamics during this energy transfer, which can depend on the detailed couplings of the inflaton to other fields.

Overall, our results highlight the fact that reheating should not be treated as a mere auxiliary phase following inflation, but rather as an integral part of inflation. Even within well-motivated and highly predictive gravitational models, uncertainties associated with reheating can significantly alter the interpretation of cosmological observables. This underlines the importance of consistently incorporating reheating effects when confronting inflationary models with data\footnote{Recent developments suggest that artificial intelligence based approaches can also play a role in systematically exploring the landscape of inflationary models and identifying viable scenarios consistent with observational data, see e.g.~\cite{Peng:2026ofs}.}. Looking ahead, improved constraints on the tensor-to-scalar ratio from forthcoming experiments such as SPIDER~\cite{SPIDER:2017xxz}, the Simons Observatory~\cite{SimonsObservatory:2018koc}, and LiteBIRD~\cite{Matsumura:2013aja} will further sharpen our ability to discriminate between competing scenarios, potentially providing valuable insights into both the inflationary dynamics and the physics of reheating.

\section*{Acknowledgements}
I would like to thank K. Tamvakis for his valuable insights and helpful discussions. This work was supported by the Estonian Research Council grants PSG1132, TARISTU24-TK10, TARISTU24-TK3, and the CoE program TK202 ``Foundations of the Universe'’. This article is based upon work from COST Action CosmoVerse CA21136, supported by COST (European Cooperation in Science and Technology).

\bibliography{references}

\end{document}